\documentclass[prd, twocolumn, nofootinbib, floatfix]{revtex4-1}

\usepackage{amsmath}
\usepackage{graphicx}
\usepackage{dcolumn}
\usepackage{bm}
\usepackage{enumerate} 
\usepackage{epsfig}
\usepackage{amssymb,latexsym,mathrsfs}
\usepackage{graphicx}
\usepackage{color}
\usepackage{hyperref}

\hypersetup{
    colorlinks=true,
    linkcolor=red,
    citecolor=blue,
} 

\newcommand{\be}{\begin{equation}}
\newcommand{\ee}{\end{equation}}
\newcommand{\bes}{\begin{equation*}}
\newcommand{\ees}{\end{equation*}}
\newcommand{\beq}{\begin{equation}}
\newcommand{\eeq}{\end{equation}}
\newcommand{\bs}{\begin{split}} 
\newcommand{\bea}{\begin{eqnarray}}
\newcommand{\eea}{\end{eqnarray}}
\newcommand{\beqa}{\begin{eqnarray}}
\newcommand{\eeqa}{\end{eqnarray}} 
\newcommand{\la}{\lesssim}

\newcommand{\om}{\Omega_m}

\newcommand{\dfom}{\dot{\rm FOM}} 

\newcommand{\dl}{\delta}
\newcommand{\tdd}{D_{\Delta t}}
\newcommand{\kext}{\kappa_{\rm ext}}
\newcommand{\gm}{\gamma'}
\newcommand{\kap}{\kappa}
\newcommand{\gam}{\gamma}
\newcommand{\sig}{\sigma}

\begin{document}

\title{Tailoring Strong Lensing Cosmographic Observations} 
\author{Eric V.\ Linder} 
\affiliation{Berkeley Center for Cosmological Physics \& Berkeley Lab, 
University of California, Berkeley, CA 94720, USA} 

\begin{abstract}
Strong lensing time delay cosmography has excellent complementarity with 
other dark energy probes, and will soon have abundant systems detected. 
We investigate two issues in the imaging and spectroscopic followup required 
to obtain the time delay distance. The first is optimization of 
spectroscopic resources. We develop a code to optimize the cosmological 
leverage under the constraint of constant spectroscopic time, and find that 
sculpting the lens system redshift distribution can deliver a 40\% 
improvement in dark energy figure of merit. The second is the role of 
systematics, correlated between different quantities of a given system or 
model errors common to all systems. We show how the levels of different 
systematics affect the cosmological parameter estimation, and derive 
guidance for the fraction of double image vs quad image systems to follow 
as a function of differing systematics between them. 
\end{abstract}

\date{\today} 

\maketitle

\section{Introduction} 

Cosmographic, or geometric, methods such as distance-redshift relations 
provide key insights into the nature of our universe. The Type Ia supernova 
luminosity distance-redshift relation revealed that the cosmic expansion is 
accelerating \cite{perl99,riess98}, with the physical cause denoted as 
dark energy. The cosmic microwave background radiation anisotropies and 
baryon acoustic oscillations in galaxy clustering are other probes that have 
at least a substantial geometric component (modulo dark energy perturbations 
or coupling to matter). Cosmic redshift drift has been recognized since the 
1960s as a potential cosmographic probe, though not yet measured 
\cite{sandage,mcvittie,fpoc,klee}. The strong gravitational lensing 
time delay distance-redshift relation was also proposed in the 1960s 
\cite{refsdal} and in the last few years has matured to the stage of being 
used as a cosmological probe \cite{suyu12,suyu13}. 

The development of strong lensing distances is a particularly interesting 
advance since unlike the standard distance-redshift relations the measured 
time delay is a dimensionful quantity, and the time delay distance is 
comprised of the ratio of three distances. This makes it sensitive to the 
Hubble constant $H_0$. Also because of this ratio the time delay distance has 
an unusual dependence on dark energy properties and has high complementarity 
with the usual distance probes \cite{lin04,lin11}. Ongoing and future 
surveys such as from the Dark Energy Survey, Large Synoptic Survey Telescope, 
and Euclid and WFIRST satellites have incorporated strong lensing time 
delays into their suite of cosmological probes. 

Here we examine two aspects of implementation of time delay distances into 
such surveys, focused on trades and optimization of the followup 
resources required to obtain a robust distance-redshift relation. In 
particular the wide field imaging surveys must be supplemented with 
spectroscopy to obtain accurate redshifts of lens and source, and to 
constrain the lens mass model. Since spectroscopy is time intensive, and 
not part of some of the planned surveys, we consider how to efficiently 
allocate the additional resources among the large numbers (1000-10000) 
of strong lens systems that will be found. 

In Sec.~\ref{sec:lensing} we review the basics of strong lensing time 
delays and the types of observations necessary to measure the time delay 
distance-redshift relation. We develop in Sec.~\ref{sec:opt} an 
optimization procedure for the cosmological leverage of the data under the 
constraint of fixed resources such as total spectroscopic time. This 
addresses questions of followup of low vs high redshift systems. The 
influence of systematic uncertainties is investigated in Sec.~\ref{sec:cov}, 
along with questions such as how to trade between different populations of 
lens systems, such as ones with double images vs quad images. We summarize and 
conclude in Sec.~\ref{sec:concl}.

\section{Measuring Time Delay Distances} \label{sec:lensing} 

The time delay distance can be thought of as the focal length of the 
lensing, and depends on the distances between observer and lens $D_l$, 
observer and source $D_s$, and lens and source $D_{ls}$. The time delay 
between two images of the source comes from the geometric path difference 
of the light propagation and from the differing gravitational potentials 
experienced. In summary, the time delay distance is 
\be 
D_{\Delta t}\equiv (1+z_l)\frac{D_l D_s}{D_{ls}}=\frac{\Delta t}{\Delta\phi} 
\ , \label{eq:dist}
\ee 
where $z_l$ is the lens redshift, $\Delta t$ is the observed time delay, 
and $\Delta\phi$ is the Fermat potential difference modeled from the 
observations such as image positions, fluxes, surface brightness, etc. 

For strong lensing time delay cosmography, the source should be a bright, 
time varying object such as an active galactic nucleus (AGN) and the lens 
is generally a foreground galaxy (as cluster lenses are harder to model). 
See \cite{araa,snowsl,ogurimars,coemous,dobke,koop,oguri07,lewisibata,futa,hjorth,sereno} 
for further details on strong lensing time delays as a cosmological probe. 

Wide field surveys such as from the Dark Energy Survey, Large Synoptic 
Survey Telescope, and Euclid and WFIRST satellites will be superb tools 
for finding 
large samples of strong lens systems. Due to their repeat observations, 
they can also monitor the image fluxes over several years to measure 
the time delay $\Delta t$. This may be supplemented with further cadenced 
observations from external programs, along the lines of the highly 
successful COSMOGRAIL program \cite{cosmograil}. 

The Fermat potential $\Delta\phi$ is constrained by the rich data of the 
images, but this works best with additional high resolution imaging, 
currently supplied by the Hubble Space Telescope, and for future surveys 
possibly by the James Webb Space Telescope (JWST) and ground based 
adaptive optics. 
For the lens part of the Fermat potential, the lens mass modeling requires 
constraint by measurement of the galaxy velocity dispersion through 
spectroscopy. This also plays a key role in breaking the mass sheet 
degeneracy \cite{treukoop,koop03,suyu13}. 
Similarly, spectroscopy obtains the redshifts of lens and 
source. (The velocity dispersion is also crucial for the possibility of 
using time delay lensing to obtain the usual angular diameter distance 
\cite{hjorth,komatsu}.) 

Thus these essential followup resources must be sought in order to derive 
the strong lensing cosmological constraints from the wide field imaging 
survey. Since these are generally external to the wide field survey, and 
require application for highly subscribed telescope time, they can become 
a limiting factor in the science return. We consider here the optimization 
of cosmological leverage given a finite followup resource. In the next 
section we present calculations specifically dealing with spectroscopy, 
but the optimization procedure is quite general. 

Since any one cosmological probe has particular degeneracies between 
parameters, we combine the strong lensing distances with cosmic microwave 
background and supernova distances. Strong lensing was shown to have great 
complementarity with these probes \cite{lin11}, and these data will exist 
at the time of the wide field surveys (indeed supernovae as distance probes 
are another component of the surveys). We adopt a Planck quality constraint 
on the distance to last scattering (0.2\%) and physical matter density 
$\om h^2$ (0.9\%). For supernovae we use a sample of the quality expected 
from ground based surveys: 150 supernovae at $z<0.1$, 900 from $z=0.1$--1, 
and 42 from $z=1$--1.7, with a statistical uncertainty of 0.15 mag and 
a systematic of $0.02(1+z)$ mag added in quadrature to each 0.1 width bin in 
redshift. 

We perform a Fisher information analysis to estimate the cosmological 
parameters of the matter density $\om$, dark energy equation of state 
present value $w_0$ and a measure of its time variation $w_a$, reduced 
Hubble constant $h$, and a nuisance parameter $\mathcal{M}$ for the 
supernova absolute magnitude. The fiducial cosmology is flat $\Lambda$CDM 
with $\om=0.3$, $h=0.7$.

\section{Optimizing Spectroscopic Followup} \label{sec:opt} 

Since spectroscopic time is restricted, and generally requires arrangements 
outside the main survey, it is advantageous to treat it as a limited 
resource and optimize its use. We consider it as a fixed quantity, and 
seek to maximize the cosmological leverage of the measured time delay 
distance given this constraint. To do so, we examine the impact of 
sculpting the redshift distribution of the lenses to be followed up. 
The spectroscopic time is dominated by measurement of the lens galaxy 
velocity dispersion. However our methodology is general and similar results 
should occur for any measurement to some given signal to noise. For 
example, one might instead optimize a resource capable of high resolution 
imaging, to map the distorted source images, such as JWST or ground based 
adaptive optics time. The principles are the same. 

For specificity, we concentrate on fixed spectroscopic time for the sample 
of lenses. To measure a redshift, or the galaxy velocity dispersion, 
requires good signal to noise data of line fluxes. Consider the following 
illustrative calculation. The signal scales with the number of photons from 
the spectral feature to be measured, hence the fluence times the exposure 
time. We will take the contribution of other sources of photons to be 
dominated by a redshift independent contribution, times the exposure time, 
so the noise, i.e.\ the fluctuations, goes as the square root of exposure 
time. This gives  
\be 
\frac{S}{N}=\frac{{\mathcal F}\,t_{\rm exp}}{N_{\rm sky}\sqrt{t_{\rm exp}}} \ . 
\ee 
We emphasize that we are presenting an illustrative methodology: true survey 
optimization will depend on many sources of noise and the survey specifications 
such as instrumental properties, scanning strategy, etc. This requires a 
full exposure time calculator and is beyond the scope of this article, but 
we will see below that our simple, heuristic approach matches some known 
results. 

Since the fluence is just the flux divided by the photon energy, we lose 
one less factor of $1+z$ than the usual flux-redshift relation, i.e.\ 
$\mathcal{F}\propto (1+z)/d_L^2$, where $d_L$ is the luminosity distance. So 
to achieve a desired constant signal to noise threshold requires the exposure 
time to vary as 
\be 
t_{\rm exp} \propto \frac{1}{\mathcal{F}^2}\propto 
\frac{d_L^4}{(1+z)^2} \propto d_A^4 (1+z)^6 \ , \label{eq:texp} 
\ee 
where $d_A=(1+z)^{-2}d_L$ is the angular diameter distance. 
At redshifts $z\approx1$--2, the angular diameter distance in our 
universe stays close to constant, and we recover the result \cite{snap} 
that spectroscopic 
exposure time becomes increasingly expensive with redshift as roughly 
$(1+z)^6$. 

At lower redshift, the slope is steeper. We will be interested 
in a range around $z\approx0.5$. However, as exposure time gets smaller, 
other noise contributions enter as well as overheads such as telescope 
slewing and detector readout time. 
Therefore we adopt a reasonable approximation that the spectroscopy cost 
goes as $t\propto (1+z)^r$ with $r=8$. We have checked that using instead 
Eq.~(\ref{eq:texp}) plus a constant overhead makes no significant difference 
in our results. We emphasize again that the key point is that for any 
$t_{\rm exp}(z)$ from an instrument's exposure time calculator, the 
optimization code described will produce results under a fixed resource 
constraint. 

The next step, given the resource constraint, is to choose the quantity to 
optimize. We take this to be the dark energy figure of merit (FOM), the 
area of a confidence contour in the dark energy equation of state plane, 
marginalized over all other parameters. The dark energy equation of state 
$w(a)=w_0+w_a(1-a)$ fits a broad range of models, and is accurate in 
recreating distances to the 0.1\% level \cite{lin03,calde}. The parameter 
$w_0$ measures the present equation of state and $w_a$ the time variation, 
with $a=1/(1+z)$ the scale factor, and 
${\rm FOM}=(\det{\rm COV}[w_0,w_a])^{-1/2}$. 
In the next section we also consider the effect of optimization on the 
determination of the Hubble constant. 

At different lens redshifts, the time delay distance has different 
sensitivities to the cosmological parameters. 
(This is true for the source redshift as well, but we fix $z_s=3z_l$ for 
simplicity; reasonable variations of this ratio have little effect on the 
cosmological sensitivity \cite{lin11}.) So the question is whether the 
extra expense of spectroscopy of higher redshift lenses overcomes their 
possibly greater leverage. 

To optimize the redshift distribution we begin with a uniform distribution 
in lens redshift (recall we are most interested in spectroscopy of the lens 
galaxy to obtain its velocity dispersion, used to constrain the lens mass 
model).  We take 25 time delay systems of 5\% precision in each bin of 
redshift width $dz=0.1$ over the range $z=0.1$--0.7, for a total of 150 
systems. This carries with it a certain total spectroscopic time, and that 
is the fixed resource constraint under which the optimization proceeds. 

This initial uniform distribution is perturbed by one system in each bin, 
one at a time, and resulting FOM is calculated. Each redshift also has a 
different time burden, and we compute the quantity 
\be 
\dfom_i=\frac{\rm FOM(perturbed)-FOM}{\Delta t_i} \ , 
\ee 
for each redshift bin $i$, where $\Delta t_i$ is the spectroscopic time 
required for a system at that redshift. The bin with the lowest $\dfom_i$, 
i.e.\ the least change in cosmological leverage, has one system subtracted 
from it. The time saved is then reallocated to the other redshift bins, 
increasing the number of systems in every other bin $j$, weighted by 
$\dfom_j$. That is, 
\be 
\Delta n_j=\Delta n_{\rm sub}\left(\frac{1+z_{\rm sub}}{1+z_j}\right)^8 
\frac{\dfom_j}{\sum_{k\ne {\rm sub}} \dfom_k} \ , \label{eq:dn} 
\ee 
where sub is the bin from which $\Delta n_{\rm sub}=1$ systems are removed. 
This formula conserves the resource, i.e.\ spectroscopic time. 

The FOM for the new distribution is computed, and the process iterates. 
The new distribution is 
perturbed, and again one system from the lowest leverage bin is removed 
(if this would cause the number in that bin to go negative, we use the 
next lowest leverage bin) and its time burden is reallocated. The iteration 
continues until convergence. As a final step we round the numbers in each 
bin to the nearest integer, but this has less than 0.3\% impact on the 
cosmology parameter estimation. This optimization method is computationally 
fast and efficient, and widely applicable to many astrophysical studies 
with constrained resources. 

\begin{figure}[htbp!]
\includegraphics[width=\columnwidth]{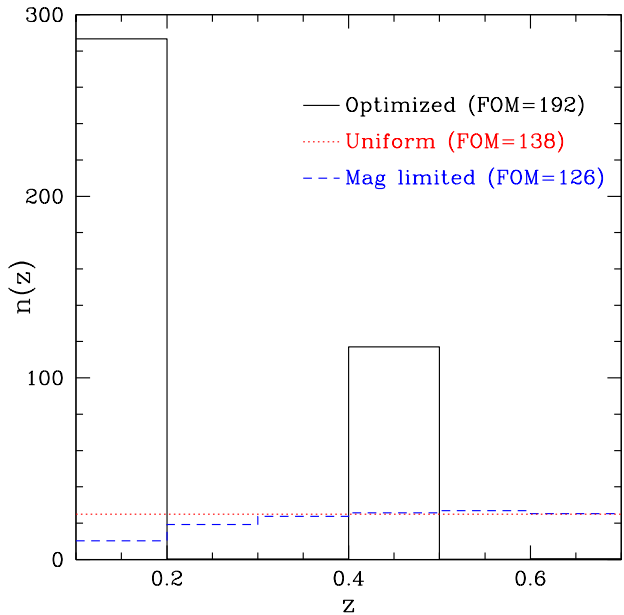} 
\caption{Histograms of the lens redshift distribution are shown for three 
cases: uniform, magnitude limited, and optimized cosmological 
leverage distributions. All three have the same fixed resource constraint 
on the total spectroscopic followup time; each is labeled with the 
resulting dark energy figure of merit. 
} 
\label{fig:nzopt} 
\end{figure}

Figure~\ref{fig:nzopt} illustrates the results. The optimization increases 
the FOM by almost 40\%, while keeping the spectroscopic time fixed. The 
optimized redshift distribution has a number of interesting properties: it 
is heavily weighted toward low redshift, with a single peak at higher, but 
not maximal redshift. Low redshift gives a decreased time burden, and still 
good cosmological leverage, especially on the Hubble constant $h$, but also 
the dark energy parameters since the source redshift extends the 
cosmological lever arm \cite{lin11}. To break covariances between parameters, 
the higher redshift bin is needed, but note it does not seek to maximize 
the range by taking the highest bin since this has the greatest time burden. 
(Note that the pioneering cosmological optimization of \cite{huttur} fixed 
the number of supernovae, not the observing time, to find a peak at the 
redshift maximum. In \cite{fhlt} both constant number resource and constant 
spectroscopic time resource with $n=6$ were studied for the supernova 
distance probe.) 
The intermediate redshift bins, and the two highest redshift bins, at 
$z=0.5$--0.6 and 0.6--0.7, are zeroed out in the optimization.

The FOM becomes a quite respectable 192, with determination of $\om$ to 
0.0035, $w_0$ to 0.061, $w_a$ to 0.22, and $h$ to $0.0030$. Each parameter 
estimation, as well as the FOM, is better than for either the uniform 
redshift distribution or a magnitude limited distribution derived from 
\cite{ogurimars} using cuts on image and lens flux and image separation 
(P.\ Marshall, S.\ Suyu private communication). We have tested reducing 
the redshift 
range to 0.1--0.4 or 0.1--0.5, and obtain the same optimized distribution, 
i.e.\ the optimum really has only the low and mid redshift bin. We also 
shifted the range to $z=0.2$--0.7 and found the optimized FOM dropped 
significantly, to 135, demonstrating the lowest bin is critical for dark 
energy as well as the Hubble constant.

\section{Influence of Systematics} \label{sec:cov} 

Any cosmological probe must deal with systematic uncertainties, 
especially for next generation surveys where abundant numbers of objects 
drive down the statistical uncertainty. We investigate two of the manifold 
aspects of the impact of systematics. Again, a detailed treatment would 
need to delve deep into survey and instrumentation properties, and is beyond 
the present scope.

\subsection{Redshift distribution revisited} 

Let us explore the effect of systematic uncertainties on the optimization 
carried out in Sec.~\ref{sec:opt}. The number of lens systems in the lowest 
redshift bin approached 300, which for individual system precision of 5\% 
implies a required control of systematic bias at the 0.3\% level. We do 
not yet know enough from current strong lensing observations and studies 
to know what is a realistic level, but strong efforts and advances in 
understanding are underway. For example, the blind data Time Delay 
Challenge (\url{http://timedelaychallenge.org}) has already achieved 
0.1--0.2\% control of time delay estimation \cite{tdc0,tdc1,hoj,amir}. 

Therefore we study the impact of various levels of systematic on the 
optimized redshift distribution and the resulting cosmological parameter 
estimation. We implement the systematic as a floor, added in quadrature 
to the statistical uncertainty, 
\be 
\frac{\sigma^2}{n_i}=\frac{\sigma^2_{\rm stat}}{n_i}+\sigma^2_{\rm sys} \ , 
\ee 
where $n_i$ is the number in redshift bin $i$ and each bin is treated 
independently. This model is commonly used 
in supernova distance-redshift relation studies \cite{klmm}. (But see the 
next subsection for an alternative approach to systematics.) 

\begin{figure}[htbp!]
\includegraphics[width=\columnwidth]{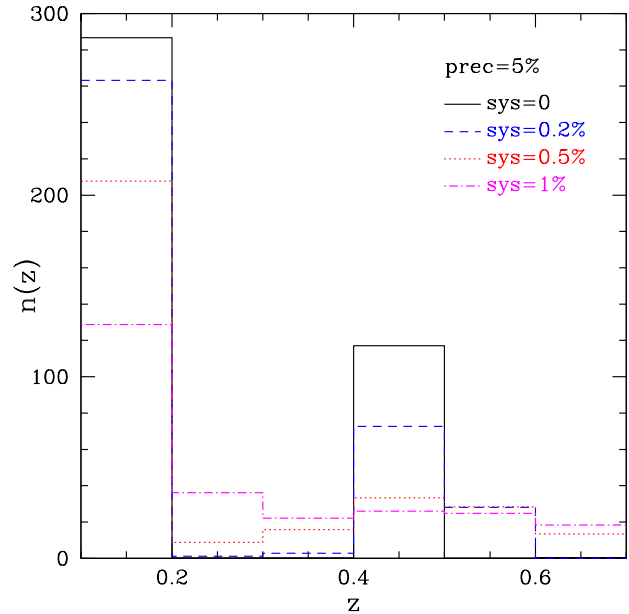} 
\caption{The resource-constrained optimization of the lens redshift 
distribution is shown as a function of coherent distance systematic floor. 
This floor prevents large numbers of systems in a redshift bin from 
improving the accuracy. 
} 
\label{fig:nzsys} 
\end{figure}

Figure~\ref{fig:nzsys} shows the optimized distributions, subject to 
the resource constraint, for different levels of systematic. As the 
systematic level increases, it is less advantageous to put a large number 
of systems in a given redshift bin and additional systems diffuse into 
neighboring bins. This fills in the intermediate redshift gap and also 
pushes some lens systems to high redshift, making the distribution closer 
to uniform.

\begin{figure}[htbp!]
\includegraphics[width=\columnwidth]{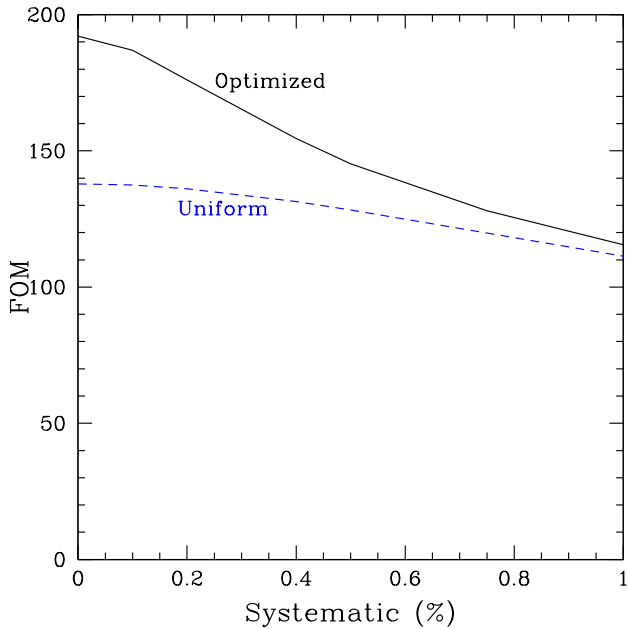} 
\caption{Dark energy figure of merit is plotted for the optimized 
cosmological leverage and uniform lens redshift distributions as a 
function of systematic floor. For high systematics, the optimized 
distribution has little extra leverage, but at low systematics the 
improvement can approach 40\%. 
} 
\label{fig:fomsys} 
\end{figure}

Both the presence of the systematic and the redistribution of the data 
away from the zero-systematic optimum lowers the FOM. Figure~\ref{fig:fomsys} 
plots the FOM vs the systematic level. At 0.2\% systematic, the FOM has 
decreased by only 8\% relative to zero systematic case, but larger 
systematics impact the cosmology more severely. For high enough levels, 
there is little difference in leverage between the optimized and uniform 
(or magnitude limited) distributions.

\begin{figure}[htbp!]
\includegraphics[width=\columnwidth]{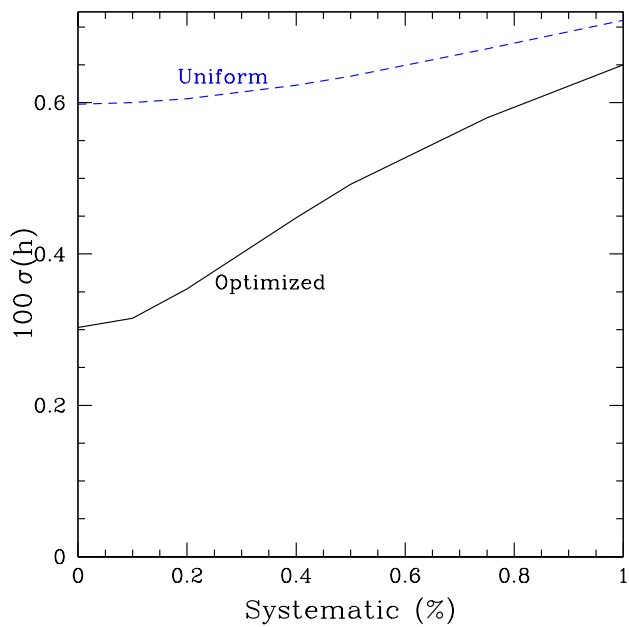} 
\caption{As Fig.~\ref{fig:fomsys} but for the Hubble constant constraint 
(note that here low values are better). At high systematics, the optimization 
still improves the results by almost 10\%, while at low systematics the gain 
is a factor 2. 
} 
\label{fig:hsys} 
\end{figure}

To make sure that optimizing for dark energy FOM also helps 
improve other cosmological parameters, we show in Fig.~\ref{fig:hsys} 
the constraint on the Hubble constant $h$. For zero systematic the 
optimization actually improves the estimation by a factor 2 (not just 
40\% as for the FOM). This is because the increased low redshift sample 
is particularly useful for the Hubble constant. Even for higher levels of 
systematics the optimization continues to give added leverage on $h$.

\subsection{Model systematics} 

Statistical uncertainties in the time delay distance arise from measurement 
imprecision, but the measurement errors can have systematic components as 
well. These take two main forms: correlated errors between elements in a 
given lensed system, and model errors that are common between different 
lens systems. These will give diagonal and offdiagonal contributions to 
the distance error matrix. Below we give an illustrative exhibition of the 
effect of such systematics; again, actual survey analysis requires a more 
detailed and sophisticated treatment but this demonstrates the main points. 

Let us write the time delay distance as 
\be
\tdd=\tdd(\Delta t,\Delta\phi(m,\vec\theta,\Delta t,v,z),\kext) \ , 
\ee
where the Fermat potential depends on image magnitudes $m$, positions 
$\vec\theta$, time delays $\Delta t$, lens and image redshifts $z$, and 
lens velocity dispersion $v$. Additional mass along the line of sight 
affects the modeling through the external convergence $\kext$. 

Ref.~\cite{suyu13} demonstrates that over the
angular range important for the images, i.e.\ near the Einstein radius of
the lens, the Fermat potential scales with the projected lens mass profile
slope $\gm$. Even when the galaxy lens profile has a multicomponent
composition \cite{sluse} of a Hernquist stellar core plus a
Navarro-Frenk-White dark matter profile,
over the angular range of interest the slope $\gm$ captures the profile
dependence. Uncertainties or misestimation in $\gm$ then lead to dispersion
or systematics in the Fermat potential and hence the distances. There is 
also a mass sheet degeneracy due to mass along the line of sight. 
Both these effects can be incorporated through \cite{suyu13} 
\be
\phi\approx \bar\phi\,(1-\kext)(\gamma'-1) \ ,
\ee
in the vicinity of the standard profile slope $\gm=2$. Note that 
systems measured to date have rms dispersion of $\la 5\%$ in $\gm$ so this
form is accurate. 

We can now write the time delay distance as
\be
\tdd=\frac{\Delta t}{\bar\phi(m,\vec\theta,\Delta t,v,z)\,
(1-\kext)(\gamma'-1)} \ . \label{eq:tdd}
\ee
The error propagation to the time delay distance, for accurate image flux and 
position measurements, is then
\be
\dl D=D_t\dl t+D_v \dl v +D_z \dl z +D_\kap \dl\kap +D_\gam \dl\gam \ ,
\ee
where we use the simplifying notation $D=\tdd$, $D_x$=$\partial D/\partial x$, 
and $\gam=\gamma'$, $\kap=\kext$, $t=\Delta t$.

A reasonable first approach, based on current data, is that the error
budget will be dominated by the time delay estimation, external convergence,
and lens mass profile and velocity dispersion. In this case the diagonal
entries in the error matrix (i.e.\ for a single lens system) would be
\bea 
C_{DD}&\approx&D^2_t \sig^2_t+D^2_v \sig^2_v+D^2_\kap \sig^2_\kap
+D^2_\gam \sig^2_\gam \nonumber \\ 
&\qquad& +D_v D_\gam\langle \dl v\,\dl\gam\rangle
+D_\kap D_\gam\langle\dl \kap\,\dl\gam\rangle \ . \label{eq:cdiag}
\eea 
This reflects the individual errors, plus the correlated errors between $v$
and $\gam$ in the lens density profile (e.g.\ whether $v$ is measured at the
appropriate place in the profile), and between $\gam$ and $\kap$ in the
profile-mass sheet degeneracy. We emphasize this is illustrative. 

The model errors enter in the off diagonal elements. Recall these correlate
two different lens systems at distances $D$ and $D'$. Here the error matrix 
gets the contribution
\bea
C_{DD'}&=& D_t D'_t \langle \dl t\,\dl t'\rangle +
D_v D'_v \langle \dl v\,\dl v'\rangle +
D_\kap D'_\kap \langle \dl \kap\,\dl \kap'\rangle \nonumber \\ 
&+& D_\gam D'_\gam \langle \dl \gam\,\dl \gam'\rangle 
+D_v D'_\gam \langle \dl v\,\dl \gam'\rangle +
D'_v D_\gam \langle \dl v'\,\dl \gam\rangle \nonumber \\ 
&+& D_\kap D'_\gam \langle \dl \kap\,\dl \gam'\rangle  +
D'_\kap D_\gam \langle \dl \kap'\,\dl \gam\rangle \ . \label{eq:coff}
\eea
Several of the derivatives can be written in a straightforward manner
using Eq.~(\ref{eq:tdd}):
\bea 
D_t&=&\frac{D}{\Delta t} \quad ;\quad D_\gam=-\frac{D}{\gam-1} 
\label{eq:deriv}\\ 
D_\kap&=&\frac{D}{1-\kext} \quad ;\quad
D_v=D_\phi\frac{\partial\phi}{\partial v}\approx -2\frac{D}{v} \ . \notag 
\eea 
The approximation sign in $D_v$ represents the result for a singular 
isothermal sphere. 

Once we are in the regime of thousands of strong lenses 
\cite{coemous,ogurimars}, the $\sqrt{N}$ statistical reduction will be 
dominated by the residual systematics.  It is crucial to identify the 
effects of these systematics on cosmological results, and where the 
greatest leverage lies in controlling them. This has significant interplay 
with our previous analysis of constrained followup resources. That is, 
we want to identify where best to concentrate the limited resources, 
e.g.\ on long time delay, quad image systems. 
One example is that double images often occur at different radii where the 
lens slope profile may vary, while quads suffer less from such a systematic. 
Quad systems may have better precision from the extra 
measurement constraints, but possibly also an increased opportunity for 
differential microlensing or varying external convergence. 

For a tractable first approach, we note that Eqs.~(\ref{eq:cdiag}), 
(\ref{eq:coff}), and (\ref{eq:deriv}) have the property that 
the uncertainties often enter into the error
covariance matrix as logarithmic fractional quantities, i.e.\ $\dl\gm/(\gm-1)$
or $\dl\kap/(1-\kap)$, times the time delay distance. We adopt the Ansatz 
that these fractional systematic uncertainties have a scaling with redshift
$s=s_\star\,[(1+z)/(1+z_\star)]^n$, where $z_\star=0.4$ is the midpoint of 
the redshift range, and a possible population dependence, where the
errors in double systems may differ from those in quad systems. Since the
fraction of doubles vs quads detected in a survey changes with redshift,
this causes a population drift in a manner similar to supernova subtype
evolution.

The overall covariance matrix will then have two entries in each redshift
bin, for doubles and quads, with distinct statistical and systematic
errors. We sum up all the statistical errors from the time delay estimation,
mass profile slope, etc.\ to give diagonal entries of
$\sig^2_{\{d,q\}}(z_i)\,D_i^2$ in the $i$th lens redshift bin. Here $D_i$ 
is the time delay distance of Eq.~(\ref{eq:dist}). The systematic errors 
$s_{\{d,q\}}(z_i)\,D_i$ also contribute to the diagonal elements and moreover 
their correlations produce offdiagonal entries. While in actual
data analysis one might not bin the data, and the error model will become more 
sophisticated over time, this approach using five parameters ($\sig_d$,
$\sig_q$, $s_d$, $s_q$, and $n$ if desired) is tractable and gives important
first indications of the effect of systematics.

We first discuss direct redshift evolution of the systematics, and then 
the influence of population drift. 

\begin{figure}[!hbt]
\begin{center}                                                                 
\includegraphics[width=\columnwidth]{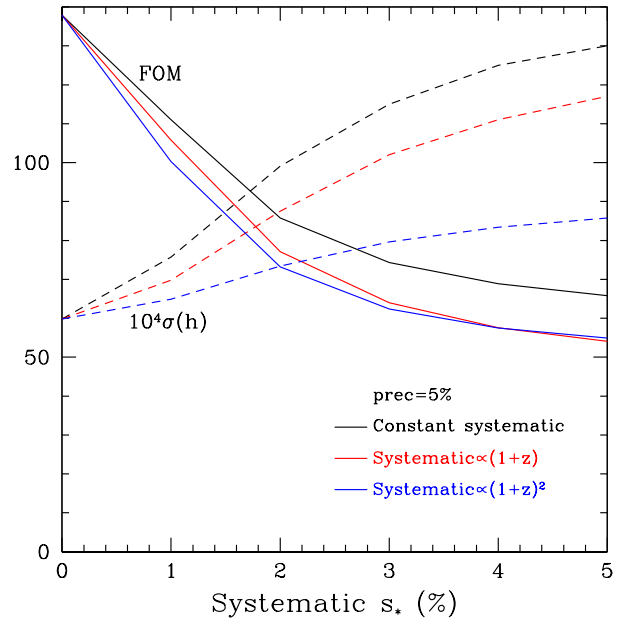} 
\caption{The dark energy figure of merit (solid curves) and
Hubble constant uncertainty (dashed curves) are shown as a function of
systematic error amplitude $s_\star$ at $z_\star=0.4$, for three different 
redshift evolutions. The systematic scales as $(1+z)^n$ for $n=0$, 1, 2. 
Note that high figure of merit is good, and low $\sigma(h)$ is good. 
} 
\label{fig:sysn}
\end{center} 
\end{figure}

Figure~\ref{fig:sysn} shows the influence of the redshift dependence of
the systematic. In order to focus on the redshift evolution of the 
systematic, we here take the uniform redshift distribution 
of Sec.~\ref{sec:opt} (despite its lower FOM), 
and treat all populations (doubles or quads) as having 
5\% statistical precision in distance and a fractional distance systematic 
$s$ per redshift bin. 
The evolution in $s$ with redshift could arise from, e.g., decreased
signal to noise, and hence more uncertain modeling, of higher redshift
lens systems (and a longer path length so greater uncertainty in the projected
mass along the line of sight).

We see that the specific systematic redshift evolution model, as opposed to 
the mere presence of the systematic, has a modest 
effect, with the variation among $n=0-2$ affecting the dark energy figure of 
merit by $\sim20\%$ and determination of the Hubble constant by $\sim50\%$ 
at the highest systematic levels. Note that the evolution does change the 
covariance between parameters; this is responsible for the $n=1$ and $n=2$ 
curves crossing at high systematic. Although $w_0$ and $w_a$ are both better 
determined in the $n=1$ case, the FOM is not, due to their altered covariance. 
As expected, a constant fractional systematic has more of an effect on $h$, 
and so less on the FOM. 

As a next step, we include separate systematics for the populations of
double and quad image systems. Again the motivation is that these have
different levels of constraints on the lens model from the observations, 
e.g.\ image positions, flux ratios, number of time delays. To focus on this
population aspect, we keep the explicit systematic amplitude independent
of redshift ($n=0$; recall we just saw that the cosmology constraints were
fairly insensitive to $n$ anyway) but incorporate population drift with
redshift.

The distribution of doubles and quads we use is the magnitude limited 
sample of Sec.~\ref{sec:opt}, based on \cite{ogurimars} (again for 
illustration, despite its lower FOM). This arises from 
the constraints that 1) image separation is larger than $1''$,
2) the quasar images are brighter than i-magnitude 20.8, and 3) the lens 
magnitude is brighter than 22. These help ensure that images and arcs
can be well resolved, time delays can be accurately measured, and the
lens velocity dispersion can be measured with reasonable use of resources.
The ratio of doubles to quads varies from roughly 9 to
4 from the low to high end of the redshift range, with statistical scatter 
in numbers included. The drift in
the proportion leads to an effective redshift evolution in the systematic
uncertainty, analogously to how population drift of supernovae subtypes
with slightly different absolute magnitudes engenders supernova magnitude
evolution (see \cite{likesys,johansys} for detailed treatment of the
propagation of this effect into cosmology constraints).

\begin{figure}[!hbt]
\begin{center}
\includegraphics[width=\columnwidth]{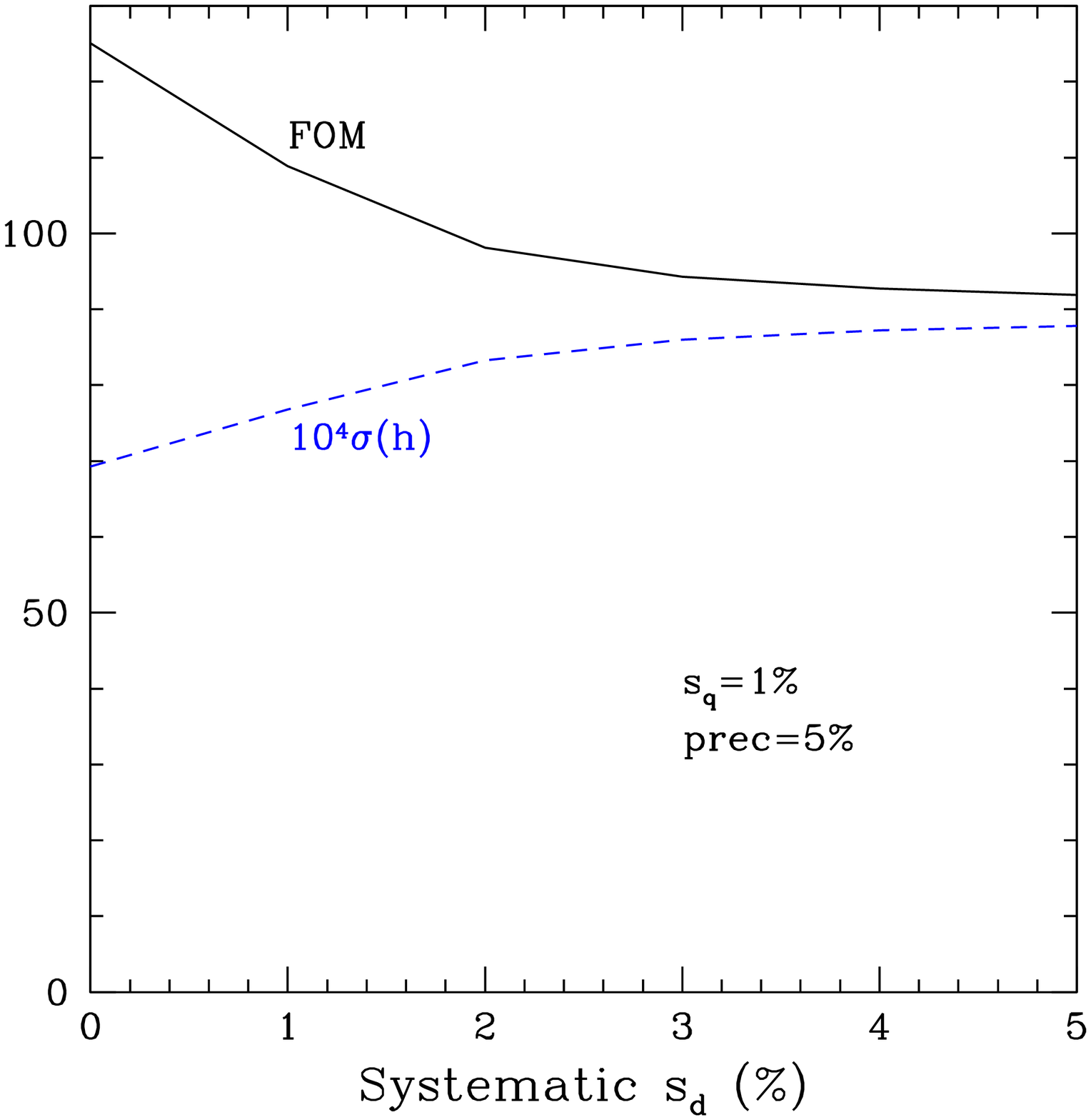} \ 
\includegraphics[width=\columnwidth]{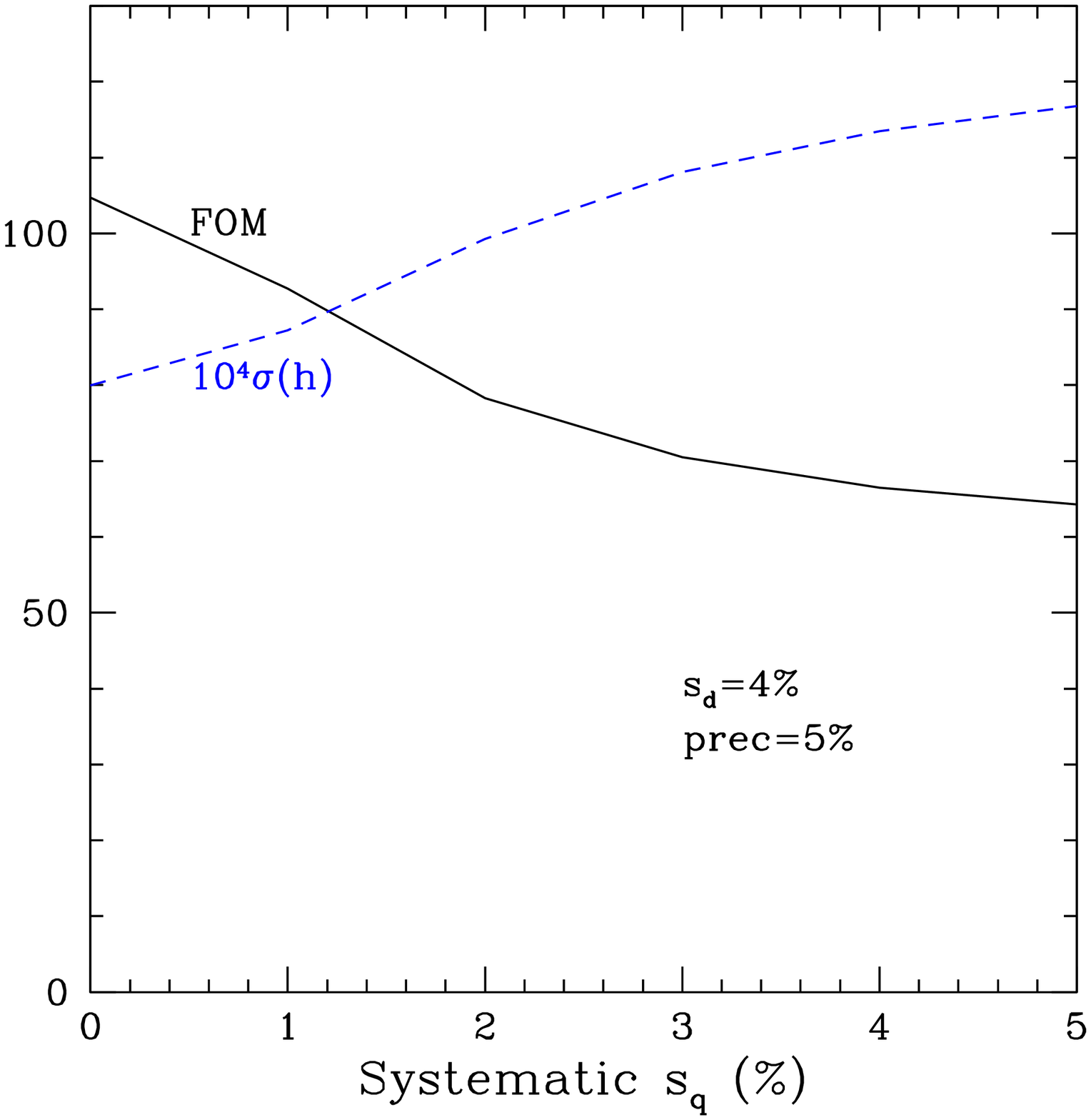} 
\caption{[Top panel] The dark energy figure of merit (solid curve) and
Hubble constant uncertainty (dashed curve) are shown as a function of
fractional systematic error $s_d$ in double image systems, for a fixed
quad image systematic $s_q=0.01$. 
[Bottom panel] As the top panel, but as a function of
$s_q$ for fixed $s_d=0.04$. 
}
\label{fig:syspop}
\end{center}
\end{figure}

Figure~\ref{fig:syspop} quantifies how population drift between the
double image and quad image systems propagates into cosmological constraints 
as the amplitude and ratio of the double and quad systematic
errors varies. The top panel fixes $s_q=1\%$ and varies $s_d$, while the 
bottom panel fixes $s_d=4\%$ and varies $s_q$. As the systematic level 
increases, the dark energy figure
of merit decreases and the uncertainty in the determination of the Hubble 
constant increases, as expected. However, when the systematic error of the
doubles exceeds their statistical uncertainty, then the information
from the doubles saturates, the quads dominate the leverage, and the FOM 
and $\sigma(h)$ level off (see top panel). This implies that systematics
in doubles have a natural ``knee'' -- defining when the limited 
observational resources have greater leverage when used to accurately
characterize the rarer quad systems. The bottom panel shows that as long
as the quad systematic is lower than its statistical uncertainty, then more
accurate measurement of quads leads to more stringent cosmological
constraints. Such an analysis can inform the optimal use of resources 
for strong lensing time delay distances as a cosmological probe.

\begin{figure}[!hbt]
\begin{center}
\includegraphics[width=\columnwidth]{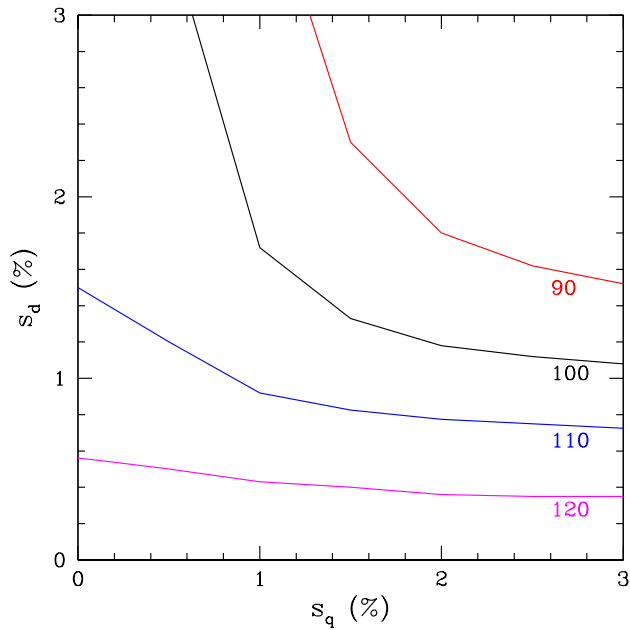} 
\caption{Contours of constant FOM are plotted as a function of the systematic 
levels in double and quad lensing systems. The zero systematics point has 
FOM=126. 
}
\label{fig:contour}
\end{center}
\end{figure}

Figure~\ref{fig:contour} depicts the results for simultaneous variation 
of the double and quad systematics, showing contours of constant FOM. 
At high systematics level, having the two contributions be comparable gives 
the best FOM. However at low systematics level we are below the quad 
statistical uncertainty. This gives two approaches for improvement: either 
improving the doubles systematic or allocating more resources to followup 
more quad systems and bring down their statistical error. From the distance 
between the contours, we see that if low quad systematics can be achieved, 
then considerable improvement in doubles systematics is required for 
significant improvement; thus following up more quad systems seems a better 
option in this case.

\section{Conclusions} \label{sec:concl} 

The strong gravitational lensing time delay distance-redshift relation is 
a geometric probe of cosmology. It has two particularly valuable 
characteristics: being dimensionful and hence sensitive to the Hubble 
constant $H_0$, and being a triple distance ratio and hence with different 
parameter degeneracies that make it highly complementary to other distance 
probes such as the cosmic microwave background or supernovae. Moreover, the 
observations and modeling are rapidly advancing, enabling it to place 
cosmological constraints of significant leverage, comparable to other methods. 

We considered the question of the followup resources needed to complement the 
forthcoming strong lensing imaging surveys that detect and monitor the lens 
systems. Since high resolution imaging or spectroscopic followup is limited 
and expensive, we optimized the lens system redshift distribution to give 
maximal cosmology leverage. The optimization code under fixed resources such 
as spectroscopic time (e.g.\ to measure the lens galaxy velocity dispersion 
to constrain the lens mass model) is computationally fast and efficient, 
with its algorithm generally applicable to many astrophysical studies and 
figures of merit. 

The sculpted distribution delivers a nearly 40\% improvement in dark energy 
figure of merit, and a factor two tighter constraint on the Hubble constant, 
than a uniform redshift distribution. Low redshift systems are found to be 
particularly preferred, and there is no need to spend followup time on 
lenses with $z>0.5$. 

Systematics enter as correlated quantities within a given lensing system, 
and as model systematics common to many systems. We examined both in an 
illustrative model that captures key aspects. A systematic uncertainty 
floor somewhat spreads out the optimal redshift distribution, but preserves 
the advantage of low redshift. We then demonstrated the effects of both a 
systematic explicitly evolving in redshift and one caused by population drift 
between different lens system types, such as double image vs quad image 
systems. If the systematic level for one population is larger than for the 
other, we can quantify by how much the followup resources are better spent 
on the more accurate population. 

As wide field surveys deliver 1000--10000 strong lensing systems, the 
issue of followup will become a key limitation, and these optimization 
tools can significantly improve the cosmological leverage. Similarly, 
as our measurement of strong lensing systems improves, the illustrative 
systematic correlation model here will become more realistic and enable 
more sophisticated trade studies regarding low vs high redshift, or 
double vs quad image systems. This will further optimize future surveys 
to use strong lensing time delay distances as a unique cosmological probe.

\acknowledgments 

I thank Alex Kim, Phil Marshall, Ramon Miquel, Sherry Suyu, and Tommaso Treu 
for helpful discussions. 
This work has been supported by DOE grant DE-SC-0007867 and the Director, 
Office of Science, Office of High Energy Physics, of the U.S.\ Department 
of Energy under Contract No.\ DE-AC02-05CH11231. 


\end{document}